\documentstyle[twocolumn,pra,aps]{revtex}
\begin{document}
\draft
\title{Error prevention scheme with two pairs of qubits}
\author{Chui-Ping Yang \thanks{%
Email address: cpyang@floquet.chem.ku.edu} and Shih-I Chu \thanks{%
Email address: sichu@ku.edu}}
\address{Department of Chemistry, University of Kansas, and Kansas Center\\
for Advanced Scientific Computing, Lawrence, Kansas 66045}
\author{Siyuan Han \thanks{%
Email address: han@ku.edu}}
\address{Department of Physics and Astronomy, University of Kansas,\\
Lawrence, Kansas 66045}
\maketitle

\begin{abstract}
A scheme is presented for protecting one-qubit quantum information against
decoherence due to a general environment and local exchange interactions.
The scheme operates essentially by distributing information over two pairs
of qubits and through error prevention procedures. In the scheme, quantum
information is encoded through a decoherece-free subspace for collective
phase errors and exchange errors affecting the qubits in pairs; leakage out
of the encoding space due to amplitude damping is reduced by quantum Zeno
effect. In addition, how to construct decoherence-free states for $n$-qubit
information against phase and exchange errors is discussed.
\end{abstract}

\pacs{PACS number{s}: 03.67.Lx, 03.65.Ta}



Quantum computing has been paid much interest in the Cavity QED, trapped ion
system, NMR system and Solid state system using nuclear spins, quantum dots,
SQUIDs, Josephson Junctions, and single-cooper-pair devices. It is realized
that one of the main obstacles in realizing a quantum computer is
decoherence resulting from the coupling of the system with environment.
Among those designed to protect information, there are theoretical proposals
for preventing quantum information against errors by using quantum Zeno
effect [1-4]. Compared with conventional error correction schemes, the
decoherence-reducing strategies based on the Zeno effects are significantly
simpler since they only require making tests on a system but no error
correction steps are needed. The most important point is that they can
reduce the number of qubits involved in the encoding of a quantum state.

Recently, using the Zeno effect, Hwang et al. [4] considered how to protect
information in an error model where phase errors are dominant but other
errors are still non-negligible. Their schemes are based on encoding
one-qubit information $\alpha \left| 0\right\rangle $ $+\beta \left|
1\right\rangle $ through a code $\left| 0_L\right\rangle =\left|
01\right\rangle $ and $\left| 1_L\right\rangle =\left| 10\right\rangle $.
Without doubt, their schemes work perfectly if there is no qubit-qubit
exchange interaction [5,6]. However, it is obvious that exchange interaction
(exchanging the qubits) turns the encoded state $\alpha \left|
01\right\rangle $ $+\beta \left| 10\right\rangle $ into $\alpha \left|
10\right\rangle $ $+\beta \left| 01\right\rangle ,$ which leads to
potentially fatal consequences (i.e, another term for bit-flipping error
appears in the resulting state (8) of Ref. [4], not only the phase errors as
mentioned there). Therefore, their schemes can not work in the presence of
exchange interaction.

In this paper, an alternative scheme is proposed for protecting one-qubit
information against decoherence due to a general environment and local
exchange interaction, based on the method of pairing qubits [7-9] and the
Zeno effect. In this scheme, the original message is encoded through two
pairs of qubits (a four-qubit encoding). The present code forms a
decoherence-free subspace (DFS) [7, 10-12] for collective phase and exchange
errors, if the following approximation applies, namely, (a) the exchange
interaction between the two pairs can be negligible (this is possible by
setting the two pairs apart, since the exchange effects generally decreases
rapidly as the qubit-qubit distance increases [6]), and (b) the two qubits
in each pair are close to each other so that each pair undergoes collective
decoherence.

Consider two separate pairs $I$ and $II$ each containing two qubits. The
four identical qubits are labeled by 1, 1$^{\prime },$ 2 and 2$^{\prime }.$
Qubits 1 and $1^{\prime }$ form the pair $I$ while qubits $2$ and 2$^{\prime
}$ constitute the other pair $II$. The two qubits in either pair are assumed
to be close to each other so that they will undergo collective decoherence.
Under the assumption that the exchange interaction between the two pairs is
small enough to be negligible, the Hamiltonian for the qubit system and the
environment is therefore of the form

\begin{equation}
H=H_S+H_B+H_{SB}+H_{EX},  \label{1}
\end{equation}
where $H_S$ and $H_B$ denote the qubit system, and the environment free
Hamiltonians, respectively; $H_{SB}$ is the interaction Hamiltonian, and the
operator $H_{EX}$ corresponds to local exchange interactions between the two
qubits in either pair. If the two pairs are physically identical, i.e, the
separation of the qubits in each pair is the same, the operator $H_{EX}$
will act simultaneously and identically on both pairs of qubits. In this
case, $H_{EX}$ acts as a collective exchange operator which has the
following form 
\begin{equation}
H_{EX}=J\left( E_{11^{\prime }}+E_{22^{\prime }}\right)  \label{2}
\end{equation}
($J$ is a constant; $E_{ij}$ is an independent exchange operator for two
identical qubits $i$ and $j,$ which has the property of $E_{ij}\left|
\epsilon _i\epsilon _j\right\rangle =\left| \epsilon _j\epsilon
_i\right\rangle $ , $\epsilon _i\in \{0,1\}$ [6]). The expressions for $%
H_S,H_{SB}$ are shown as follows 
\begin{eqnarray}
H_S &=&\epsilon _0\left( \sigma _I^z+\sigma _{II}^z\right) ,  \nonumber \\
H_{SB} &=&\lambda _1^z\sigma _I^z\otimes V_z+\lambda _1^{+}\sigma
_I^{+}\otimes V_{+}+\lambda _1^{-}\sigma _I^{-}\otimes V_{-}  \nonumber
\label{5} \\
&&\ +\lambda _2^z\sigma _{II}^z\otimes V_z^{\prime }+\lambda _2^{+}\sigma
_{II}^{+}\otimes V_{+}^{\prime }+\lambda _2^{-}\sigma _{II}^{-}\otimes
V_{-}^{\prime }.  \label{3}
\end{eqnarray}
Here, $\sigma _I^j=\sigma _1^j+\sigma _{1^{\prime }}^j,$ $\sigma
_{II}^j=\sigma _2^j+\sigma _{2^{\prime }}^j$ $\left( j=z,+,-\right) ;$ $%
\sigma _i^j$ is Pauli spin operators of the qubit $i$; $V_j$ and $%
V_j^{\prime }$ are the environment operators coupled to these degrees of
freedom. This interaction Hamiltonian $H_{SB}$ applies to the situation: the
qubits inside each pair undergoes collective decoherence while the two pairs
undergo independent decoherence for the case of different $V_j$ and $%
V_j^{\prime }$ or imperfect collective decoherence for the case of the same $%
V_j$ and $V_j^{\prime }$.

Suppose that qubit 1 is the original information carrier, which is initially
in an arbitrary unknown state $\left| \psi \right\rangle =\alpha \left|
0\right\rangle +\beta \left| 1\right\rangle .$ The encoding is shown as
follows 
\begin{equation}
\left| \psi \right\rangle _{enc}=\alpha \left| 0\right\rangle _L+\beta
\left| 1\right\rangle _L,  \label{4}
\end{equation}
where 
\begin{eqnarray}
\left| 0_L\right\rangle &=&\left( \left| 01\right\rangle +\left|
10\right\rangle \right) _{11^{\prime }}\left( \left| 01\right\rangle -\left|
10\right\rangle \right) _{22^{\prime }},  \nonumber \\
\left| 1_L\right\rangle &=&\left( \left| 01\right\rangle -\left|
10\right\rangle \right) _{11^{\prime }}\left( \left| 01\right\rangle +\left|
10\right\rangle \right) _{22^{\prime }}.  \label{5}
\end{eqnarray}
This encoding will protect the state (4) against collective phase errors
taking place at either pair or both, since the qubits 1$1^{\prime }$ and 22$%
^{\prime }$ are paired up in the DF state (decoherence-free state)
combinations $\left| 01\right\rangle $ and $\left| 10\right\rangle $.
Moreover, it is obvious that the collective exchange operator (2) has the
property of $H_{EX}\left| 0\right\rangle _L=\left( E_{11^{\prime
}}+E_{22^{\prime }}\right) \left| 0\right\rangle _L=0$ and $H_{EX}\left|
1\right\rangle _L=\left( E_{11^{\prime }}+E_{22^{\prime }}\right) \left|
1\right\rangle _L=0$, which shows that the independent exchange error for
each pair cancels each other due to the cooperative action between the local
exchange interaction in one pair and the local exchange interaction in the
other pair, i.e., the code also forms a DFS for exchange errors.

Suppose that the environment is initially in the state $\left| \psi _b\left(
0\right) \right\rangle .$ During a finite time $T_0,$ perform $N$ times
tests. In a short period of time $T_0/N,$ under the Hamiltonian (1), the
encoded state (4) will evolve into 
\begin{eqnarray}
\left| \psi \left( T_0/N\right) \right\rangle &\approx &\left[ 1-iH\left(
T_0/N\right) \right] \left| \psi \right\rangle _{enc}\otimes \left| \psi
_b\left( 0\right) \right\rangle  \nonumber \\
\ &=&\left[ \alpha \left( \left| 01\right\rangle +\left| 10\right\rangle
\right) _{11^{\prime }}\left( \left| 01\right\rangle -\left| 10\right\rangle
\right) _{22^{\prime }}\right.  \nonumber \\
&&\ \ +\left. \beta \left( \left| 01\right\rangle -\left| 10\right\rangle
\right) _{11^{\prime }}\left( \left| 01\right\rangle +\left| 10\right\rangle
\right) _{22^{\prime }}\right]  \nonumber \\
&&\otimes \left[ 1-iH_B\left( T_0/N\right) \right] \left| \psi _b\left(
0\right) \right\rangle  \nonumber \\
&&\ \ -i\left( T_0/N\right) \left| 11\right\rangle _{11^{\prime }}\left(
\left| 01\right\rangle -\left| 10\right\rangle \right) _{22^{\prime }} 
\nonumber \\
&&\otimes \lambda _1^{+}\alpha V_{+}\left| \psi _b\left( 0\right)
\right\rangle  \nonumber \\
&&\ \ -i\left( T_0/N\right) \left| 00\right\rangle _{11^{\prime }}\left(
\left| 01\right\rangle -\left| 10\right\rangle \right) _{22^{\prime }} 
\nonumber \\
&&\otimes \lambda _1^{-}\alpha V_{-}\left| \psi _b\left( 0\right)
\right\rangle  \nonumber \\
&&\ \ -i\left( T_0/N\right) \left( \left| 01\right\rangle -\left|
10\right\rangle \right) _{11^{\prime }}\left| 11\right\rangle _{22^{\prime }}
\nonumber \\
&&\otimes \lambda _2^{+}\beta V_{+}^{\prime }\left| \psi _b\left( 0\right)
\right\rangle  \nonumber \\
&&\ \ -i\left( T_0/N\right) \left( \left| 01\right\rangle -\left|
10\right\rangle \right) _{11^{\prime }}\left| 00\right\rangle _{22^{\prime }}
\nonumber  \label{6} \\
&&\otimes \lambda _2^{-}\beta V_{-}^{\prime }\left| \psi _b\left( 0\right)
\right\rangle .  \label{6}
\end{eqnarray}

Eq. (6) shows that after the evolution for a short time $T_0/N$, if one
performs a measurement in succession to determine whether the four-qubit
system has left the encoding space spanned by (5), the probability for
getting the result ``out of the encoding space'' is of the order of $1/N^2$,
and therefore, the probability of obtaining such an outcome during the time $%
T_0$ is proportional to $1/N$. Taking $N$, the number of tests during the
time $T_0$, large enough one can decrease the probability of such an error
below any desired level. On the other hand, after the evolution of time $%
T_0/N$ the state inside the encoding space remains the same as the initial
encoded state, and the probability of obtaining such an outcome during the
time $T_0$ is proportional to $1-O\left( 1/N\right) $.

The required projection can be performed in two steps. The first step is to
prepare a test qubit (labeled by $t)$ in the state $\left| 0\right\rangle $
, make it interact with each of the two qubits in the first pair $I$
consecutively by a joint operation $C_{1t}C_{1^{\prime }t}$ and then perform
a measurement on the test qubit. The measurement outcome $\left|
1\right\rangle $ projects the whole system onto the state 
\begin{eqnarray}
\left| \psi \left( T_0/N\right) \right\rangle ^{\prime } &=&a\left[ \alpha
\left( \left| 01\right\rangle +\left| 10\right\rangle \right) _{11^{\prime
}}\left( \left| 01\right\rangle -\left| 10\right\rangle \right) _{22^{\prime
}}\right.  \nonumber \\
&&\ +\left. \beta \left( \left| 01\right\rangle -\left| 10\right\rangle
\right) _{11^{\prime }}\left( \left| 01\right\rangle +\left| 10\right\rangle
\right) _{22^{\prime }}\right]  \nonumber \\
&&\ +b\left( \left| 01\right\rangle -\left| 10\right\rangle \right)
_{11^{\prime }}\left| 11\right\rangle _{22^{\prime }}  \nonumber \\
&&\ +c\left( \left| 01\right\rangle -\left| 10\right\rangle \right)
_{11^{\prime }}\left| 00\right\rangle _{22^{\prime }},  \label{7}
\end{eqnarray}
while $\left| 0\right\rangle $ corresponds to the projection onto the state 
\begin{eqnarray}
\left| \psi \left( T_0/N\right) \right\rangle ^{\prime \prime } &=&d\left|
11\right\rangle _{11^{\prime }}\left( \left| 01\right\rangle -\left|
10\right\rangle \right) _{22^{\prime }}  \nonumber  \label{8} \\
&&+e\left| 00\right\rangle _{11^{\prime }}\left( \left| 01\right\rangle
-\left| 10\right\rangle \right) _{22^{\prime }}.  \label{8}
\end{eqnarray}
Under the condition of large $N$, the effects of the state (8), which is
outside the encoding space, can be negligible. Thus, after this test step,
the state of the four qubits and the environment will be in the state (7).

The second step follows the same procedure as described above. One needs to
have the test qubit (in the zero state) interact with each of the two qubits
in the second pair $II$ by a joint operation $C_{2t}C_{2^{\prime }t}$ and
then make a measurement on the test qubit. From Eq. (7) one can see that the
measurement outcome $\left| 0\right\rangle $ projects the whole system onto
the state 
\begin{equation}
b\left( \left| 01\right\rangle -\left| 10\right\rangle \right) _{11^{\prime
}}\left| 11\right\rangle _{22^{\prime }}+c\left( \left| 01\right\rangle
-\left| 10\right\rangle \right) _{11^{\prime }}\left| 00\right\rangle
_{22^{\prime }},  \label{9}
\end{equation}
which is the wrong state out of the encoding space, and again the effects of
this state (9) can be neglected if one performs his tests frequently enough;
on the other hand, if the test qubit is measured in the state $\left|
1\right\rangle ,$ the four qubits will remain in the original encoded state
(4). Thus, after the time $T_0,$ the final state for the whole system will
be given by 
\begin{equation}
\left| \psi \left( T_0\right) \right\rangle \approx \left| \psi
\right\rangle _{enc}\otimes \left| \widetilde{\psi }_b\right\rangle ,
\label{10}
\end{equation}
where $\left| \widetilde{\psi }_b\right\rangle $ is the state of the
environment. It is clear that no errors in the encoded state (4) occur after
overall time evolution. Thus, one can protect one-qubit information against
decoherence without any other error correction.

The present scheme works by the Zeno effect, thus it can deal only with
``slow'' noise. The characteristic time of the noise coupling has to be
larger than the time interval between the projection measurements. These
limitations are also required by other error prevention schemes based on the
quantum Zeno effects [1-4].

One might envision to use Vaidman's code [1] 
\begin{eqnarray}
\left| 0_L\right\rangle &=&\left( \left| 00\right\rangle +\left|
11\right\rangle \right) \left( \left| 00\right\rangle +\left|
11\right\rangle \right) ,  \nonumber  \label{1} \\
\left| 1_L\right\rangle &=&\left( \left| 00\right\rangle -\left|
11\right\rangle \right) \left( \left| 00\right\rangle -\left|
11\right\rangle \right)  \label{11}
\end{eqnarray}
to accomplish the goal. As long as the exchange interaction between the left
two qubits and the right two qubits is small enough to be negligible, this
code also forms a DFS for exchange errors. It is noted that the code (11)
works for the case of each qubit undergoing independent decoherence, i.e.,
the left or the right two qubits in (11) do not need to be set close. In
this sense, the scheme of Vaidman et al. is better than the present scheme
since it has a less strict condition. However, as was argued by Vaidman [1],
after a short-time evolution, the test qubit has to interact with $all$ $%
four $ $physical$ $qubits$ of the system consecutively to detect phase
errors, besides interacting with every two physical qubits of the system to
distinguish bit-flip errors. In contrast, since the present code forms a DFS
for collective phase errors, no phase errors occur and thus no such a step
for detecting phase errors is required. As shown above, the present scheme
only needs to detect bit-flip errors, by a test qubit interacting with two
qubits for each test step. Therefore, the present error prevention
procedures are much simpler.

Duan and Guo [2] have shown that one-qubit information can be protected
against decoherence due to a general environment with only two qubits and
the assistance of an external driving field. The present scheme, however,
focuses on how to protect one-qubit information without using an external
driving field and how to reduce decoherence arising from qubit-qubit
exchange interaction.

Another point may need to be made here. If there is no exchange interaction,
and if a general environment affects qubits independently , $\left|
0_L\right\rangle $ and $\left| 1_L\right\rangle $ in (4) could be the
logical zero and one of the five-qubit [13] or seven-qubit codes [14]; or
they could be the logical zero and one of the four-qubit code [15].

In what follows, our purpose is to show how to construct DF states for $n$%
-qubit quantum information against collective phase and exchange errors. The
general state of $n$ qubits is expressed as 
\begin{equation}
\left| \psi \right\rangle =\sum_{\left\{ i_l\right\} }c_{\left\{ i_l\right\}
}\left| \left\{ i_l\right\} \right\rangle ,  \label{12}
\end{equation}
where $\left| \left\{ i_l\right\} \right\rangle $ represents a computational
basis state $\left| i_1\right\rangle \otimes \left| i_2\right\rangle \otimes
\cdot \cdot \cdot \otimes \left| i_n\right\rangle $ with $i_l=0$ or $1.$ The
state (12) is encoded into the following state of $n+2$ pairs 
\begin{equation}
\left| \psi \right\rangle _{enc}=\sum_{\left\{ i_l\right\} }c_{\left\{
i_l\right\} }\left| \left\{ i_l\right\} \right\rangle _L,  \label{13}
\end{equation}
here, 
\begin{eqnarray}
\left| \left\{ i_l\right\} \right\rangle _L
&=&\prod\limits_{k=1}^{n+2}\left| j_{kk^{\prime }}\right\rangle  \nonumber
\label{14} \\
&=&\left| j_{11^{\prime }}\right\rangle \otimes \left| j_{22^{\prime
}}\right\rangle \otimes \cdot \cdot \cdot \otimes \left| j_{\left(
n+2\right) \left( n+2\right) ^{\prime }}\right\rangle .  \label{14}
\end{eqnarray}
In (14), $\left| j_{kk^{\prime }}\right\rangle $ indicates the encoded zero
or one of the $k$th pair, which is given by 
\begin{eqnarray}
\left| 0_{kk^{\prime }}\right\rangle &\rightarrow &\frac 12\left( \left|
01\right\rangle +\left| 10\right\rangle \right) _{kk^{\prime }},  \nonumber
\\
\left| 1_{kk^{\prime }}\right\rangle &\rightarrow &\frac 12\left( \left|
01\right\rangle -\left| 10\right\rangle \right) _{kk^{\prime }},  \label{15}
\end{eqnarray}
where $kk^{\prime }$ represents the two qubits in the $k$th pair. Clearly,
such an encoding (15) on each pair ensures that the encoded state (13) is a
DF state for collective phase errors if the two qubits in each pair are
close to each other.

Assume that the separation of the two qubits in each pair is the same and
that the exchange interaction between any two pairs is negligible. Thus the
collective exchange operator $H_{EX}$ is 
\begin{equation}
H_{EX}=J\sum\limits_{k=1}^{n+2}E_{kk^{\prime }}.  \label{16}
\end{equation}
It is worth noting that not all the DF states for phase errors are DF states
for exchange errors, since exchanging the two qubits in each pair will make $%
\left| 0_{kk^{\prime }}\right\rangle \rightarrow \left| 0_{kk^{\prime
}}\right\rangle $ while $\left| 1_{kk^{\prime }}\right\rangle \rightarrow
-\left| 1_{kk^{\prime }}\right\rangle $ (for the latter, there is a
phase-flip error). However, one still can expect that the encoded state (13)
is a DF state for exchange errors, through an appropriate encoding on each
pair and making the encoded state (13) be an eigenstate of the collective
exchange operator (16).

In order to have the encoded state (13) to be an eigenstate of the
collective exchange operator (16), one needs to make each logical state in
the encoded state (13) be an eigenstate of the collective exchange operator
(16) with the same eigenvalue. In general, for $n+2$ pairs of qubits, one
can construct $C_{n+2}^m$ orthogonal states. Each of them takes the form
(14) and all of them are eigenstates of the collective exchange operator
(16) with the same eigenvalue $J\left( n-2m+2\right) $ (where $m=1,2,\cdot
\cdot \cdot ,\frac{n+1}2$ for odd $n$ and $m=1,2,\cdot \cdot \cdot ,\frac n2%
+1$ for even $n$). It is easy to see that (a) $C_{n+2}^m$ reaches maximum
when $m=\frac{n+1}2$ for odd $n$ or $m=\frac n2+1$ for even $n,$ and (b)
such a maximum satisfies the relation $n<\log _2C_{n+2}^m<n+1$. The point
(a) means that in the case when each orthogonal state is an eigenstate of
the collective exchange operator (16) with the same eigenvalue $J$ for odd $%
n $ or 0 for even $n,$ the number of such orthogonal states is maximal; the
point (b) implies that all these orthogonal states, as logical states $%
\{\left| \left\{ i_l\right\} \right\rangle _L\},$ are sufficient to encode $%
n $ logical qubits. Thus, $n+2$ pairs of qubits are sufficient to encode an
arbitrary state of $n$ qubits into a DF state. For a large $n,$ the
efficiency of the encoding is approximately 1/2. On the other hand, it is
easy to show that $n+1$ pairs of qubits are not sufficient to do above.

It is interesting to note that for some kinds of entangled state of $n$
(distant) qubits, the DF states for collective phase and exchange errors can
be obtained by pairing each entangled qubit with an ancilla qubit and
applying local operation on each pair. For example, consider the following
entangled state 
\begin{eqnarray}
\left| \Psi \right\rangle ^{\left( 1\right) } &=&\alpha _0\left|
0\right\rangle _{12...\left( n-1\right) }\left| 1\right\rangle _n  \nonumber
\label{17} \\
&&+\sum\limits_{i=1}^{n-1}\alpha _i\left| n-2,1\right\rangle _{12...\left(
n-1\right) }^{\left( i\right) }\left| 0\right\rangle _n,  \label{17}
\end{eqnarray}
where the number of entangled qubits $n\geq 3,$ and $\left|
n-2,1\right\rangle _{12...\left( n-1\right) }^{\left( i\right) }$ denotes
the $i$th computational basis state of the $n-1$ entangled qubits involving $%
n-2$ zeros and 1 ones. In the case of $\left| \alpha _0\right| =\left|
\alpha _i\right| =\frac 1{\sqrt{n}},$ the state (17) are known as the
entangled W states [16]. If each entangled qubit is paired with an ancilla
qubit and then the two orthogonal states $\left| 0\right\rangle $ and $%
\left| 1\right\rangle $ of the original $k$th entangled qubit are encoded
into the logical zero $\left| 0_{kk^{\prime }}\right\rangle $ and one $%
\left| 1_{kk^{\prime }}\right\rangle $ in (15) respectively, one can see
that the resulting encoded state for the state (17) is an eigenstate of the
collective exchange operator $H_{EX}=J\sum\limits_{k=1}^nE_{kk^{\prime }}$
with an eigenvalue $\left( n-2\right) J$, i.e., the encoded state is a DF
state for exchange errors; and it is also a DF state for collective phase
errors if collective decoherence holds for each pair.

In addition, entangled state of the form 
\begin{equation}
\left| \Psi \right\rangle ^{\left( 2\right) }=\alpha \left|
i_1i_2...i_n\right\rangle +\beta \left| \overline{i}_1\overline{i}_2...%
\overline{i}_n\right\rangle  \label{18}
\end{equation}
(which, in the case $\left| \alpha \right| =\left| \beta \right| =\frac 1{%
\sqrt{2}},$ are known as entangled GHZ states [17]) are widely used in
information processing. Here, the $i_j$ are ones or zeros and $\overline{i}%
_j $ are their complements. By pairing each entangled qubit with an ancilla
qubit and performing the same encoding on each pair as above, one can see
that the two components $\left| i_1i_2...i_n\right\rangle _L$ and $\left| 
\overline{i}_1\overline{i}_2...\overline{i}_n\right\rangle _L$ in the
encoded state $\left| \Psi \right\rangle _{enc}^{\left( 2\right) }=\alpha
\left| i_1i_2...i_n\right\rangle _L+\beta \left| \overline{i}_1\overline{i}%
_2...\overline{i}_n\right\rangle _L$ are eigenstates of the collective
exchange operator $H_{EX}=\sum\limits_{k=1}^nJ_{kk^{\prime }}E_{kk^{\prime
}} $ with an eigenvalue $\sum\limits_{k=1}^n\left( -1\right)
^{i_k}J_{kk^{\prime }}$ for $\left| i_1i_2...i_n\right\rangle _L$ while $%
\sum\limits_{k=1}^n\left( -1\right) ^{\overline{i}_k}J_{kk^{\prime }}$ for $%
\left| \overline{i}_1\overline{i}_2...\overline{i}_n\right\rangle _L$. It is
easy to show that after evolving for time $t$, the $n$ pairs of qubits will
be in the state 
\begin{equation}
\alpha \left| i_1i_2...i_n\right\rangle _L+e^{i\varphi }\beta \left| 
\overline{i}_1\overline{i}_2...\overline{i}_n\right\rangle _L,  \label{19}
\end{equation}
where $\varphi =t\sum\limits_{k=1}^n\left[ \left( -1\right) ^{i_k}-\left(
-1\right) ^{\overline{i}_k}\right] J_{kk^{\prime }}$. This accumulated phase
factor in the final state might not be significant for the states (18) in
some applications. Furthermore, if (a) the number of the originally
entangled qubits is even, (b) the number of 1's is the same as that of 0's
in each of the two basis states of equation (18), (c) $J_{kk^{\prime }}=J$,
the phase factor $\varphi $ will be zero. In this case, the encoded state is
perfectly protected against collective phase and exchange errors during the
time evolution.

So far, a three-qubit error correction code [18-20] and a two-qubit error
prevention code [1,3], which protect one-qubit information against phase
damping and exchange errors, have been proposed. Compared with these
schemes, the present method has the advantage of not requiring error
correction or error detection. Moreover, compared with the schemes [18-20],
the present method requires less qubit resource in protecting the entangled
states (17) and (18), or in protecting $n$-qubit information ($n\geq 5$).
Thus, the present method is more efficient, although one has to have the two
qubits in each pair close to each other and all the pairs to be well
separated.

Finally, according to the above description, for each pair: leakage out of
the encoding subspace spanned by (15), due to amplitude damping, can be
suppressed by frequent tests on each pair. Thus, for a general environment, $%
n$-qubit information or above $n$-qubit entangled states can also be
protected by encoding them into above DF states and plus the Zeno effect.

In conclusion, we have presented an error prevention scheme for protecting
one-qubit information against decoherence due to a general environment and
local exchange interactions. As shown above, the present error prevention
procedures are relatively simple. We have discussed how to construct DF
states for $n$-qubit information against collective phase and exchange
errors. Moreover, we have shown that certain kinds of important entangled
states of $n$ (distant) qubits can be protected, by pairing each entangled
qubit with only one ancilla qubit and applying only local operations on each
pair.

CPY is very grateful to L. M. Duan and J. Gea-Banacloche for helpful
discussions. This work was partially supported by US National Science
Foundation (EIA-0082499).

\end{document}